\documentclass[cameraready]{Interspeech}

\title{Deriving Benchmarking Datasets from Long-Form Recordings: \\ Challenges and Opportunities}

\author[affiliation={1}, orcid=0009-0008-3193-6983, correspondingauthor]{Kaveri K.}{Sheth}
\author[affiliation={1}, orcid=0000-0000-0000-0000]{Lawrence}{Borst}
\author[affiliation={1}, orcid=0009-0009-2158-597X]{Tarek}{Kunze}
\author[affiliation={2}, orcid=0000-0002-6005-9368]{Marvin}{Lavechin}
\author[affiliation={3}, orcid=0000-0002-0537-0946]{Okko}{Räsänen}
\author[affiliation={1}, orcid=0000-0001-9580-4500]{Sho}{Tsuji}
\author[affiliation={1}, orcid=0000-0003-0509-5494]{Loann}{Peurey}
\author[affiliation={1}]{Alix}{Bourrée}
\author[affiliation={1}, orcid=0000-0003-2979-4556]{Alejandrina}{Cristia}

\address{
    $^1$ LAAC, LSCP, DEC, ENS, EHESS, CNRS, PSL University, Paris, France \\
    $^2$ Laboratoire d'Informatique et Systèmes, Université Aix-Marseille, CNRS, France \\
    $^3$ Signal Processing Research Centre, Tampere University, Finland
}

\email{ksheth@ens.psl.eu}


\usepackage{comment}

\usepackage{makecell}
\usepackage{float}
\usepackage{tikz}
\usetikzlibrary{arrows.meta, positioning}

\newcommand{\class}[1]{\small \texttt{#1}}

\begin{document}

\maketitle

\begin{abstract}
Long-form recordings (LFRs) of child-centered audio are ecologically valid sources for studying early language development, but three problems limit their use. First, LFR corpora are collected across sites with heterogeneous formats and consent structures, making cross-corpus use non-trivial. Second, without standardized benchmarks, assessing whether tools generalize across languages and conditions is hard. Third, ML workflows rarely respect privacy constraints governing sensitive child speech. This paper presents a framework addressing all three: a standardized collection of 27 child-centered datasets built with open-source tools (S1); a replicable pipeline for four speech-processing benchmarks (S2); and ELSI, a role-based ecosystem embedding ethical governance into the ML workflow (S3). We demonstrate the framework via a voice type classification case study and show the three solutions are mutually dependent. 
    
\end{abstract}

\section{Introduction}

Long-form recordings (LFRs) of child-centered audio, typically obtained via wearable microphones worn by young children throughout a full day, provide unmatched ecological validity for studying language input, production, and acquisition \cite{casillas2024daylong, cychosz2020longform, peurey25_interspeech}. However, three problems have prevented LFR corpora from being fully leveraged for speech tool development and evaluation.

\textbf{Problem 1: Cross-corpus heterogeneity makes joint use non-trivial.} LFR corpora have been collected by independent research teams worldwide, each with its own annotation format, metadata conventions, directory structure, and participant consent framework. A team wishing to train a classifier on several corpora, or evaluate a model across different linguistic communities, must currently work around these differences for each new project. While inconvenient, inconsistent annotation mappings can also introduce errors, and mismatched consent frameworks can create legal and ethical exposure. The practical consequence is that most tools are trained and evaluated on a single corpus or a small subset, limiting their generalizability.

\textbf{Problem 2: The absence of shared benchmarks slows development.} Without standardized evaluation sets spanning diverse languages, child ages, and recording contexts, it is difficult to compare models across papers, identify pain points, or track progress over time. To our knowledge, no shared benchmarks currently exist for child-centered speech processing comparable to those available for adult speech. Importantly, this is not just a matter of data being unavailable: even for corpora with human annotations, deriving a reusable benchmark requires resolving the heterogeneity problem described above first. A benchmark built on a single corpus or English-dominant data will also give an overly optimistic picture of how well tools generalize across the linguistic and cultural diversity of LFR research contexts.

\textbf{Problem 3: Standard ML workflows are not designed for sensitive child speech data.} LFRs inevitably capture privacy-sensitive moments of people's everyday lives and incidentally record third-party individuals who may have not consented to participate to the data collection \cite{cychosz2020longform}. Platforms such as HomeBank \cite{vandam_homebank_2016} and Databrary \cite{gilmore2018practical} have established tiered access controls for raw audio, but they do not govern the downstream ML workflow (e.g. model training, evaluation, and the distribution of derived outputs) where additional privacy risks occur. Without governance that covers the full pipeline, researchers may inadvertently expose participant data — for example, through model artifacts that memorize training examples, shared evaluation results, or benchmark splits from which individual speakers can be identified.

These three problems are interdependent. Solving heterogeneity (S1) is a prerequisite for deriving meaningful benchmarks (S2), because a benchmark built on annotations that have not been standardized across corpora cannot serve as a reliable evaluation standard. Producing and distributing those benchmarks (S2) requires governance (S3), because most contributing corpora are under access restrictions that prohibit unrestricted distribution of audio or annotation segments. And governance (S3) is only tractable when datasets share a common organizational structure (S1), enabling automated auditing of data provenance and consent. A solution addressing only one or two of these problems leaves the others unresolved.

This paper presents a framework that addresses all three jointly, through solutions that build on one another (Figure~\ref{fig:ecosystem}):
\begin{enumerate}
    \item A collection of \textbf{27} child-centered datasets \textbf{standardized} using DataLad and ChildProject, spanning diverse languages and child-rearing contexts (Section~\ref{sec:contrib1});
    \item A replicable pipeline for deriving four \textbf{benchmark datasets} from this collection, demonstrated through a voice type classification case study (Section~\ref{sec:contrib2});
    \item \textbf{ELSI}, a role-based access ecosystem that embeds ethical governance directly into the ML research workflow (Section~\ref{sec:contrib3}).
\end{enumerate}

\section{Background}

\textbf{Annotation cost and partial solutions.} 
Because exhaustive human annotation is infeasible, the field has developed automated tools to segment audio and derive metrics from full recordings. The most widely used is the LENA system~\cite{gilkerson2008lena}, a commercial tool that classifies audio into broad categories such as child vocalizations and adult speech and produces summary counts per recording. More recently, neural approaches such as the Voice Type Classifier~\cite{babyhubert_charlot2025} have been developed to perform similar speaker-type segmentation with greater accuracy. Human annotation is reserved for shorter clips sampled from the full recording, and can capture richer information than automated tools — for example, the addressee of an utterance (whether speech is directed at the child or another adult), phonetic features, or word-level transcriptions. This motivated the development of the DARCLE Annotation scheme~\cite{casillas2017new}, which establishes a standardized process for human annotation of key speech features in child-centered recordings. Similarly, the so-called ACLEW project introduced the ACLEW-DAS (Daylong Audio Segmentation) protocol, which has enabled somewhat comparable annotations across studies\footnote{\url{https://osf.io/b2jep/}}. These annotation schemes typically cover one or more of the following domains: (a) voice type classification and segmentation; (b) vocalization types; (c) addressee identification; and (d) orthographic transcription. These efforts have produced useful annotations but remain scattered across individually managed corpora.

\textbf{Privacy and data sharing.} Established repositories such as HomeBank~\cite{vandam_homebank_2016} and Databrary~\cite{gilmore2018practical} have enabled raw audio to be restricted and shared only under institutional authorization. This in turn has enabled broader use in the research community while protecting participants. HomeBank supports sharing of derived outputs such as transcripts across corpora, but neither platform provides infrastructure for governing the downstream ML workflow, including model training, evaluation, and the distribution of benchmark splits.

\textbf{Cross-dataset standardization.} The ChildProject Python package for LFR data management~\cite{childproject_gautheron2023} addresses organizational heterogeneity by enforcing a consistent directory structure and metadata schema across corpora. DataLad adds git-based version control for annotations and git-annex for large file management, enabling nested dataset structures that support reproducibility. These tools have been validated within individual projects, but they have not yet been applied at scale across multi-corpus collections of LFR data.

The framework described in this paper integrates these existing tools into a unified pipeline, adding a governance layer (ELSI) that makes the entire process operable across institutions and access tiers.

\section{Solution 1: Standardizing (many) corpora}
\label{sec:contrib1}

Aggregating data across independently collected corpora is a common challenge in fields where no single dataset is large or diverse enough to support generalizable models. Doing so reliably requires at minimum: a shared organizational schema, consistent metadata conventions, version control to track dataset changes over time, and large-file management for audio data. Without these, cross-corpus pipelines are difficult to reproduce.

Several data repositories exist for child language research. CHILDES~\cite{macwhinney2001childes} and the California Language Archive~\cite{skilton2021ticuna} provide access to transcripts and recordings, and HomeBank~\cite{vandam_homebank_2016} and Databrary~\cite{gilmore2018practical} offer tiered access controls for sensitive audio. However, none of these platforms govern the downstream research workflow, enforce a common annotation schema, or provide tools for cross-corpus aggregation. In practice, researchers who wish to pool data across sites must resolve format inconsistencies manually which is time-consuming, error-prone, and rarely documented in enough detail to be reproduced.

Over four years, we have assembled a collection of child-centered audio datasets from collaborating researchers worldwide and standardized them using two open-source packages: DataLad\footnote{\url{datalad.org}} for versioning and large-file management, and ChildProject\footnote{\url{childproject.readthedocs.io}} for organizational structure and metadata~\cite{childproject_gautheron2023}.

\textbf{Organizational structure.} Each corpus follows a common schema with metadata at three levels: child (e.g., ID, date of birth), recording (e.g., recording ID, child ID, date), and annotation. The annotation-level metadata, generated automatically by ChildProject, maps each annotation file to its corresponding audio segment, since human annotations of LFRs typically cover only sampled audio sections rather than full recordings.

\textbf{Annotation sets.} Each corpus may contain multiple annotation sets. For instance, a \textbf{vtc} set holds the output of the automated Voice Type Classifier~\cite{babyhubert_charlot2025}, an \textbf{its} set holds the output of the LENA system~\cite{gilkerson2008lena}, and an \textbf{eaf} set holds human annotations created in ELAN. Each human annotation set is accompanied by a metadata file documenting the annotation method, the types of information annotated (e.g., speaker type, addressee, transcription), and sampling parameters (e.g., whether clips were sampled periodically, randomly, or targeting regions of high child vocalization). This metadata removes the need for users to inspect each corpus's raw files to determine what was annotated and how, centralizing that information in a single queryable index.

\textbf{Dataset overview.} Table~\ref{tab:corpora-overview} summarizes the 27 datasets we have standardized and pooled together that include at least one human annotation set (out of 40 total in our collection). Six are publicly accessible; of these, five were extracted from CHILDES~\cite{macwhinney2001childes} and one from the California Language Archive~\cite{skilton2021ticuna}. The remaining 21 require access controls due to the sensitive nature of the recordings.

The collection spans 18+ languages across 14 countries, with key children aged 0--7 years. The cross-linguistic and cross-cultural nature is a deliberate priority. Recent analyses have shown that CHILDES overrepresents monolinguals and highly educated families~\cite{scaff2025demographic}, and that model performance on child speech degrades substantially outside the conditions represented in training data~\cite{blasi2022systematic}. Restricting the collection to public datasets would mean working almost exclusively with English-language, WEIRD-population recordings.

\begin{table*}[t]
    \centering
    \caption{Overview of corpora included in the derived benchmark datasets, split into public (top) and non-public long-form corpora (bottom). \emph{N clips}: number of annotated clips; \emph{Dur}: total duration of audio that has been human-annotated in minutes; \emph{VTC/Addressee/VCM/Trans. utts}: number of annotated utterances per task. VTC: Voice Type Classification; VCM: vocal maturity; Trans: Transcription}
    \label{tab:corpora-overview}
    \resizebox{\textwidth}{!}{
    \begin{tabular}{llrrrrrrrr}
        \toprule
        Corpus & Language & Location & N clips & Dur (min) & VTC utts & Addr. utts & VCM utts & Trans. utts \\
        \midrule
        \multicolumn{9}{l}{\textit{Public corpora}} \\
        \midrule
        forrester~\cite{forrester_2002} & English & USA & 31 & 2,124 & 12,992 & -- & -- & 13,197 \\
        providence~\cite{providence_2006} & English & USA & 364 & 62,789 & 364,101 & -- & -- & 368,009 \\
        soderstrom~\cite{soderstrom_acoustical_2008} & English & Canada & 57 & 4,679 & 33,555 & 5,478 & -- & 33,679 \\
        thomas~\cite{thomas_2009} & English & UK & 377 & 73,532 & 552,410 & 354 & -- & 590,264 \\
        tsay~\cite{tsay_2007_construction} & Minnan & Taiwan & 245 & 24,626 & 140,328 & -- & -- & 326,975 \\
        ticuna2018~\cite{ticuna_skilton_2019} & Ticuna & Peru & 81 & 162 & 381 & -- & -- & -- \\
        \midrule
        \textbf{Total} & 3 languages & 5 countries & \textbf{1,155} & \textbf{167,912} & \textbf{1,103,767} & \textbf{5,832} & \textbf{--} & \textbf{1,332,124} \\
        \midrule
        \multicolumn{9}{l}{\textit{Non-public corpora (long-form)}} \\
        \midrule
        babylogger-vs-lena-data~\cite{babylogger_lena_2021} & French & France & 48 & 20 & 860 & -- & -- & -- \\
        bergelson~\cite{bergelson2017seedlings} & English & USA & 311 & 1,224 & 11,811 & 9,202 & 3,573 & 12,775 \\
        cougar~\cite{vandam2018cougar} & English & USA & 236 & 2,360 & 51,278 & 37,893 & 22,753 & 52,262 \\
        fausey-trio~\cite{fausey2018trio} & English & USA & 491 & 246 & 1,105 & -- & -- & -- \\
        israel-natovich~\cite{israel_haifa_natovich} & Hebrew & Israel & 60 & 15 & 5,005 & -- & -- & -- \\
        japanese-tsuji2024~\cite{japanese_tsuji_2024} & Japanese & Japan & 13 & 780 & 1,155 & -- & -- & -- \\
        lucid~\cite{lucid_rowland_2025} & English (UK) & UK & 310 & 1,220 & 14,225 & 10,243 & 5,459 & 15,703 \\
        lyon~\cite{lyon_homebank_2016} & French & France & 2,769 & 4,647 & 52,326 & 21,051 & 25,473 & 34,787 \\
        oshikoto2025-petrovic~\cite{oshikoto_petrovic_2025_namibia} & Haillom, Khewdam & Namibia & 46 & 92 & 1,080 & -- & -- & -- \\
        phonSES~\cite{phonses_cristia_2021} & French & France & 65 & 130 & 751 & 453 & -- & -- \\
        png2016~\cite{casillas_png2016} & Yélî-Dnye & Papua New Guinea & 216 & 768 & 19,064 & 14,099 & 4,382 & -- \\
        png2019~\cite{png2019_cristia} & Yélî-Dnye (and others) & Papua New Guinea & 24 & 48 & 832 & -- & -- & -- \\
        quechua~\cite{quechua_cychosz_homebank_2018} & Spanish, Quechua & Bolivia & 108 & 216 & 1,689 & -- & -- & -- \\
        solomon~\cite{cassar2025-solomon} & Many & Solomon Islands & 946 & 1,139 & 12,817 & -- & 8,496 & -- \\
        timor-leste2022~\cite{timor_leste_22_baranov} & Mambai, Kairui, Tetun-Terik, Tutun-Prasa, +15 more & Timor-Leste & 161 & 322 & 2,670 & -- & -- & -- \\
        timor-leste2024 & Mambai, Kairui, Tetun-Terik, Tutun-Prasa, +15 more & Timor-Leste & 104 & 208 & 1,578 & -- & -- & -- \\
        tseltal2015~\cite{tseltal_casillas_homebank_2017} & Tseltal & Mexico & 213 & 812 & 11,471 & 7,567 & 3,478 & 11,469 \\
        tsimane2017~\cite{tsimane2017_scaff} & Tsimane & Bolivia & 1,298 & 1,510 & 14,951 & 3,718 & 4,643 & 7,452 \\
        vanuatu~\cite{vanuatu_cristia_2023} & Vanuatu & Vanuatu & 533 & 490 & 5,118 & 177 & 7,315 & -- \\
        warlaumont~\cite{warlaumont_homebank_2024} & English, Spanish, German, Sahaptin & USA & 267 & 1,048 & 10,539 & 7,537 & 3,789 & 11,326 \\
        winnipeg~\cite{winnipeg_soderstrom_homebank_2016} & English & Canada & 243 & 954 & 9,550 & 8,694 & 3,252 & 11,949 \\
        \midrule
        \textbf{Total} & 18+ languages & 14 countries & \textbf{8,462} & \textbf{18,249} & \textbf{232,106} & \textbf{120,634} & \textbf{92,613} & \textbf{159,953} \\
        \bottomrule
    \end{tabular}
    }
\end{table*}

\section{Solution 2: Benchmarking machine learning algorithms}
\label{sec:contrib2}

The standardized collection described in Section~\ref{sec:contrib1} is what makes this solution possible: because all corpora share a common schema and queryable metadata, annotation sets can be identified and aggregated programmatically rather than by inspecting each corpus individually. As our second solution, we exploit this infrastructure to derive four benchmark datasets using a replicable DataLad-based pipeline. Because corpus datasets are nested as inputs, the pipeline can be re-run whenever a new corpus is added, an annotation set is updated, or an error is corrected, without disrupting projects that have already built on a prior version.

Selection of annotation sets was performed manually, guided by the metadata described in Section~\ref{sec:contrib1}, which significantly reduced the effort required to identify suitable sets while also revealing cases where automated selection would have been insufficient --- for instance, where time-stamp accuracy varied substantially between annotation sets within the same corpus. Each annotation set was then mapped to a common label schema per task before aggregation.

\subsection{Task Descriptions}

\textbf{Voice type classification (VTC).} Annotators labeled the onset and offset of vocalizations attributed to four speaker types: the key child (the child wearing the recorder), other children, female adults, and male adults. This segmentation is a prerequisite for most downstream analyses of LFRs. All 27 corpora contribute to this benchmark. Speaker labels were mapped to the four target types where possible; labels that could not be resolved (e.g., ``friend'' or ``cousin'' in some CHILDES corpora) were excluded. The multi-label designation also enabled handling of overlapping speech.

\textbf{Addressee classification.} Thirteen corpora contain utterance-level annotations of who the speaker is addressing. Labels were mapped to three categories available across all contributing corpora: target-child-directed speech, adult-directed speech, and other-directed speech. This benchmark can support both classification (given a segmented utterance, predict its addressee) and diarization (identify child-directed regions in continuous audio) settings. A 2017 ComParE challenge~\cite{schuller2017interspeech} on related material found the task to be highly challenging; the present benchmark constitutes an invitation to revisit it with more diverse data and stronger models.

\textbf{Vocal maturity (VCM) classification.} Eleven corpora contain utterance-level labels of the key child's vocalization type: crying, laughing, canonical babble, or non-canonical babble. Building on the ACLEW annotation protocol~\cite{soderstrom2021developing} and a 2019 ComParE challenge~\cite{schuller2019interspeech}, this task has seen recent progress with pre-trained model approaches~\cite{zhang2025employing}. Crucially, no public corpus in our collection contributes to this benchmark; all data comes from restricted LFRs, underscoring the value of the tiered access framework described in Section~\ref{sec:contrib3}.

\textbf{Orthographic transcription.} Thirteen corpora yield intelligible, lexical speech with orthographic transcriptions, suitable for automatic speech recognition evaluation or language input analysis. This is the one task for which public corpus data (primarily from CHILDES-derived corpora) has more transcription data than private corpora.

All the four benchmarks use child-disjoint splits, meaning that all recordings of a given child are assigned to the same partition, preventing a model from being trained on one recording of a child and evaluated on another. To make the benchmark available for the community, we created a replicable pipeline \footnote{https://github.com/LAAC-LSCP/benchmarking-dataset-factory} for obtaining access to the datasets and preparing the data for the benchmark.

\subsection{Case Study: Voice Type Classification}

To illustrate the utility of the benchmarks, we train a voice type classifier on the derived VTC dataset and compare it to the state-of-the-art VTC~2.0 model~\cite{babyhubert_charlot2025}. VTC~2.0 uses BabyHuBERT representations passed to four independent binary classification heads (one per speaker type), enabling multi-label prediction. We adopt the same architecture and fine-tune it on our derived dataset---first on the public corpora only, then on the full collection---training with AdamW~\cite{loshchilov2018decoupled} (batch size 256 and learning rate $1\times10^{-5}$) until convergence.

Table~\ref{tab:vtc-retrain} reports F1 scores on the hold-out test set of our benchmark. The model retrained on public data alone underperforms VTC~2.0 substantially (44.4\% vs.~65.1\% average F1), reflecting the limited size and distributional range of the public subset. The results on the private dataset are able to match and even surpass the results of the state-of-the-art model on some classes (73.4\% vs.~70.0\% F1 on \class{KCHI} and 68.8\% vs 65.1\% F1 on \class{MAL}). These results demonstrate that access to the full collection---made possible by the governance framework described in Section~\ref{sec:contrib3}---is not merely convenient but necessary for competitive performance across the variety of world's languages that are of interest in LFR-based research on early language development.

\begin{table}[ht]
    \centering
    \fontsize{8pt}{8pt}\selectfont
    \addtolength{\tabcolsep}{-0.25em}
    \caption{F1-scores (\%) obtained on the hold-out set by the state-of-the-art VTC-2.0~\cite{babyhubert_charlot2025} model and compare it to the vtc model re-trained on the public and private benchmarking corpuses. Best model performances are shown in \textbf{bold}, second best are \underline{underlined}.}
    \label{tab:vtc-retrain}
    \begin{tabular}{@{}lccccc@{}}
        \toprule
        & \multicolumn{5}{c}{F1-scores (\%) $\uparrow$} \\
        Model & \class{KCHI} & \class{OCH} & \class{MAL} & \class{FEM} & Ave. \\
        \midrule
        VTC-2.0~\cite{babyhubert_charlot2025}  & \underline{70.0} & \textbf{50.9} & \underline{65.1} & \textbf{74.3} & \textbf{65.1} \\
        VTC-2.0 - retrain public & 48.0 & 26.8 & 56.3 & 42.4 & 44.4\\
        VTC-2.0 - retrain private & \textbf{73.4} & \underline{34.2} & \textbf{68.8} & \underline{72.2} & \underline{62.2}\\
        \midrule
        Human 2 & 79.7 & 60.4 & 67.6 & 71.5 & 69.8 \\
        \bottomrule
    \end{tabular}
\vspace*{-\baselineskip}
\end{table}

\section{Solution 3: Protecting participants' privacy while advancing science}
\label{sec:contrib3}
The benchmarks described in Section~\ref{sec:contrib2} are only useful if they can be shared. Without systematic governance, the collection risks either over-restriction, which limits scientific progress, or under-restriction, which compromises participant privacy. Most LFR data cannot be made fully public due to its sensitivity, yet full restriction would undermine the goal of shared benchmarks. What is needed is a framework that delivers data to appropriate researchers at a level of access matched to use-case sensitivity while preserving participant privacy. Tiered access platforms such as HomeBank and Databrary provide important precedents for raw audio, but they do not support the full pipeline from corpus to model to derived metric. Our solution, ELSI (ExELang Legacy Support Interface), addresses this gap.

\textbf{Design rationale.} The key insight motivating ELSI is that different research activities require different levels of data sensitivity. Training a voice type classifier requires raw audio; evaluating it on a benchmark requires only audio segments with labels; analyzing the output of a trained classifier requires only derived metrics (e.g., proportion of child-directed speech per recording). A governance structure that treats all of these activities as equivalent imposes unnecessary restrictions on most users. ELSI addresses this by tying access permissions to the type of data needed — so a researcher who only needs aggregate metrics is not subject to the same restrictions as one who needs raw audio.

\begin{figure}[h]
    \centering
    \includegraphics[width=\columnwidth]{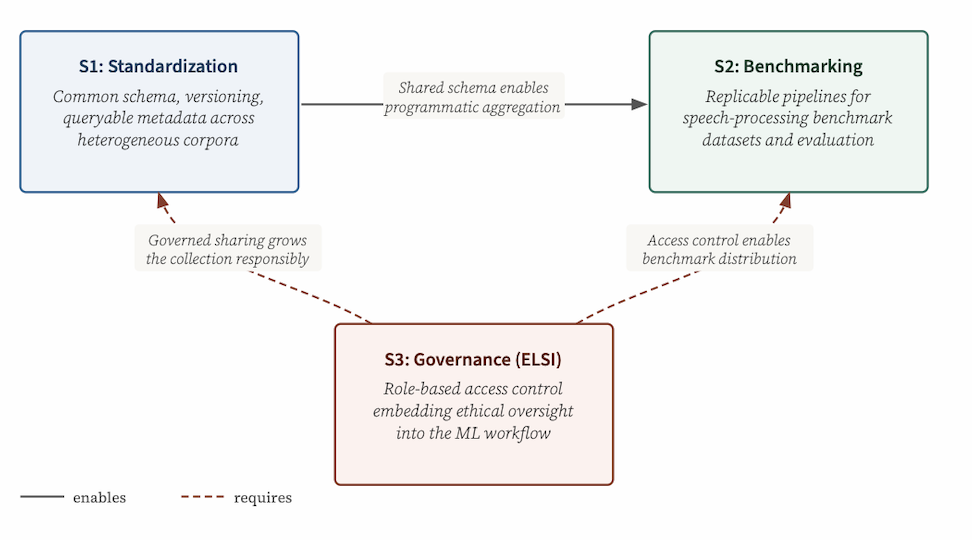}
    \caption{Interdependence of the three solutions. Standardization (S1) is a prerequisite for benchmarking (S2): a shared organizational schema and queryable metadata make cross-corpus aggregation tractable. Governance (S3) enables both: responsible data sharing expands the collection, and access control makes benchmark distribution possible. Together, the three solutions form a mutually reinforcing framework for open, reproducible, and ethically grounded child speech research.}
    \label{fig:ecosystem}
\end{figure}

\textbf{Role-based access.} ELSI operates through three roles (Figure~\ref{fig:ecosystem}). \textbf{Custodians} (ExELang Consortium administrators) control access to raw audio, grant user permissions, and ensure compliance with each corpus's consent and ethics protocols. \textbf{Tool Creators} hold custodian-approved access to raw audio and are responsible for developing, training, and validating machine learning models, which are versioned and contributed back to the ecosystem. \textbf{Analysts} work exclusively with derived metrics and de-identified outputs such as automated annotations without direct access to raw audio. 

\textbf{Connection to Solutions 1 and 2.} ELSI builds directly on the infrastructure from Section~\ref{sec:contrib1}: standardized metadata makes it possible to automatically trace where each piece of data came from and what consent covers it, and DataLad versioning ensures that analyses can be linked to specific dataset versions regardless of subsequent updates. ChildProject ensures that the included data, including any new datasets, follow the standards of the infrastructure. The benchmarks from Section~\ref{sec:contrib2} serve as the interface between tool creators and analysts. Tool creators can validate their models against benchmark splits before releasing them, and analysts can then apply those versioned models to produce the derived outputs they work with. This creates a continuous improvement cycle: analysts' findings surface new challenges, which guide further model development by tool creators, which produces better tools for analysts, all within a governance structure that keeps raw audio protected throughout.

\section{Discussion}

The VTC experiment illustrates the paper's central argument concretely. Retraining on public data alone yields substantially lower performance than the state of the art (44.4\% vs. 65.1\% average F1), because the public subset is both small and linguistically narrow. Retraining on private corpora recovers much of this gap (62.2\%), but requires a governance structure that data collectors can trust. The entire pipeline depends on each component being in place.

Broadly, the framework generalizes beyond child speech. Any domain collecting sensitive large-scale wearable data faces analogous tensions between data utility and participant protection. Open-source standardization, derived-output benchmarks, and role-based governance represent a usable path forward.

\clearpage
\section{Acknowledgments}
This work received institutional support from the Agence Nationale de la Recherche (ANR-17-EURE-0017 et ANR-10-IDEX-0001-02), and PSL (Université Paris Sciences \& Lettres). KS, LB, TK, AB, LP, ST, and AC were supported in part by the J. S. McDonnell Foundation Understanding Human Cognition Scholar Award and the European Research Council (ERC) under the European Union’s Horizon 2020 research and innovation programme (ExELang, Grant agreement No. 101001095). OR was funded by ERC grant (EveryContext, 101229121). ML acknowledges funding from Simons Foundation International (034070-00033). Views and opinions expressed are those of the authors only and do not necessarily reflect those of the European Union or the European Research Council. Neither the European Union nor the granting authority can be held responsible for them.

\section{Use of Generative AI Disclosure}
Generative AI tools (e.g., Claude) were used solely for language editing and stylistic polishing of text, including grammar and clarity improvements. All scientific content, methodology, experiments, analyses, and conclusions were conceived, verified, and approved by the authors.

\bibliographystyle{IEEEtran}
\bibliography{refs}

\end{document}